%Paper: cond-mat/9511076
%From: tellez@stat.th.u-psud.fr (Gabriel Tellez A.)
%Date: Wed, 15 Nov 1995 16:40:34 GMT

\magnification=1200
\font\boldmath=cmbsy10
\textfont2=\boldmath
\mathchardef\mynabla="0235
\font\boldgreek=cmmib10
\textfont9=\boldgreek

\mathchardef\mysigma="091B
\def\rrprime{{\bf r},{\bf r}'}
\def\correl#1#2{<#1 ({\bf r}) #2 ({\bf r}')>^{\rm T}}
\def\phiphi{\correl{\Phi}{\Phi}}
\def\piw{{\pi\over 2W}}
\def\sigmasigma{\correl{\sigma}{\sigma}}

\headline{\ifnum\pageno=1 \nopagenumbers
\else \hss\number \pageno \fi}

\overfullrule=0pt

\footline={\hfil}
\baselineskip=10pt

\def\R{ {\rm R \kern -.31cm I \kern .15cm}}
\def\C{ {\rm C \kern -.15cm \vrule width.5pt \kern .12cm}}
\def\Z{ {\rm Z \kern -.27cm \angle \kern .02cm}}
\def\N{ {\rm N \kern -.26cm \vrule width.4pt \kern .10cm}}
\def\1{{\rm 1\mskip-4.5mu l} }
\vbox to 1,5truecm{}
\centerline{\bf The Ideal Conductor Limit}

 \bigskip

\bigskip
\centerline{{\bf B. Jancovici}\footnote{$^{\dagger}$}{E-mail address~:
janco@stat.th.u-psud.fr} {\bf and G. T\'ellez}\footnote{$^{\ddagger}$}{E-mail
address~:
tellez@stat.th.u-psud.fr}} \medskip \centerline{Laboratoire de Physique
Th\'eorique et
Hautes Energies\footnote{*}{Laboratoire associ\'e au Centre National de la
Recherche
Scientifique - URA D0063}} \centerline{Universit\'e de Paris-Sud, b\^atiment
211, 91405 Orsay
Cedex, France}   \medskip

\bigskip\bigskip
\baselineskip=20pt
\noindent
${\bf Abstract}$ \par
This paper compares two methods of statistical mechanics used to study a
classical
Coulomb system $S$ near an ideal conductor $C$. The first method consists in
neglecting
the thermal fluctuations in the conductor $C$ and constrains the electric
potential to
be constant on it. In the second method the conductor $C$ is considered as a
conducting Coulomb system the charge correlation length of which goes to zero.
It has
been noticed in the past, in particular cases, that the two methods yield the
same
results for the particle densities and correlations in $S$. It
is shown that this is true in general for the quantities which depend only on
the
degrees of freedom of $S$, but that some other quantities, especially the
electric
potential correlations and the stress tensor, are different in the two
approaches. In
spite of this the two methods give the same electric forces exerted on $S$.

\vbox to 1cm{}
%\noindent {\bf KEYWORDS}~:

\vbox to 1.5cm{}

\noindent LPTHE Orsay 95-71 \par
\noindent November 1995
\vfill\supereject \noindent {\bf 1. Introduction} \vskip 5mm
In the equilibrium statistical mechanics of classical Coulomb systems (for
instance
electrolytes), sometimes one is led to assume that some wall is an ideal
conductor (for
instance for mimicking an electrode). Two methods have been used to deal with a
classical (i.e. non quantum) Coulomb system $S$ near an ideal conductor $C$.
The
first one is to consider from the beginning that the conductor $C$ is ideal and
take this into account by constraining the electric potential to be constant on
$C$
[1-3]. The second method is to treat the conductor $C$ as a genuine Coulomb
system with
a microscopic structure and take the limit of zero correlation length [4, 5]~;
indeed, in
that limit, the statistical average of the charge density on the conductor $C$
becomes a
surface charge density of zero thickness, a characteristic feature of ideal
conductors.
\par

Both methods gave the same results for some quantities in the Coulomb system
$S$,
such as the particle densities, and even the fluctuations of these densities as
described by the particle correlation functions. The reason for this agreement
about some fluctuations is not obvious. In the first approach, there are no
fluctuations inside the conductor $C$. In the second approach, there are
thermal
fluctuations inside the conductor $C$~; for instance, the potential-potential
correlation function has a universal simple form [6] (in 3 dimensions, $k_BT$
divided by
the distance) for distances large compared to the microscopic scale and these
correlations do not disappear as the zero correlation length limit is taken.
However, this seems to have no influence on some quantities in the Coulomb
system
$S$. Nevertheless we shall show that some other quantities in $S$ - for
instance
the electric potential correlations - are different depending on the method
used
to compute them. \par

The aim of the present paper is to discuss the relationship between the two
approaches. We first treat a simple two-dimensional example in section~2, where
we consider two parallel lines~; one is the ideal conductor $C$ and the other
one
is the Coulomb system $S$. The two methods are worked out to find the electric
potential and charge correlations and the stress tensor, and we compare the
results. In section~3, we treat the general case, look at quantities such as
the
partition function, the correlations and the stress tensor, and again we
compare the
results obtained through the two methods. \par

The reader may either look first at section~2, or go directly to section~3.

\vskip 5mm
\noindent {\bf 2. A simple two-dimensional example}
\vskip 5mm
\noindent {\it 2.1. The model}
\vskip 5mm
For simplicity, we consider a system of ``restricted dimension'' [7]. The
Coulomb
interaction between two point-charges $q$ and $q'$ at a distance $r$ from each
other has
the two-dimensional form $-qq'\ln \ r$, but the particles are constrained to
live on
one-dimensional lines. In the plane $xOy$, the line $y = 0$ is the conductor
$C$ at
zero potential, while the Coulomb system $S$ lives on the line $y = W$ ($W >
0$). \par

We consider two cases~: either the conductor $C$ is an ideal one from the
beginning, or
it is the high-density limit of a Coulomb system (in that limit, the
microscopic scale
goes to zero). We compute the potential and charge correlations and the average
stress
tensor in each case. These quantities can be obtained by a
ma\-cros\-copic approach [6], using linear response theory and a conducting
behaviour
assumption. Alternatively, exact microscopic results can be derived in a
special model.
\par

A position ${\bf r} = (x, y)$ is conveniently represented by the complex number
$z = x
+ iy$.

\vskip 5mm
\noindent {\it 2.2. The ideal conductor approach}
\vskip 5mm
The line $y = 0$ is assumed to be a grounded ideal conductor $C$, in that sense
that the
electric potential is constrained to vanish on that line. Thus, in the region
$y,y' > 0$,
the potential at $z'$ due to a unit point-charge at $z$ is
$$
G({\bf r},{\bf r}')=-\ln \left| {z-z' \over z-\bar{z}'} \right|
\eqno(2.1)
$$
\noindent which does vanish on the line $y = 0$. A Coulomb system $S$ lies on
the line
$y = W$. This model has been already studied [7]. Under the assumption that $S$
is a
conductor, a ma\-cros\-copic approach [6] allows to compute the correlation
function for
the electric potential $\Phi$ at two points ${\bf r}$ and ${\bf r}'$ (provided
$W$ is
macroscopic and ${\bf r}$ and ${\bf r}'$ are at macroscopic distances from the
line $y
= W$). These correlations are ($<...>^{\rm T}$ means a truncated statistical
average and
$\beta$ is the inverse temperature)
$$
\eqalignno{
\beta \phiphi&=-\ln\left|{z-z'\over z-\bar{z}'}\ {{\sinh\piw (z-\bar{z}')}
\over {\sinh\piw (z-z')}}\right|
\cr
&\hbox{if}\quad (y,y')\in ]0,W[^2 \ \ \ , &(2.2{\rm a})
\cr\cr
\beta \phiphi&=-\ln\left|{z-z'\over z-\bar{z}'}\right|
 \cr
&\hbox{if} \quad(y,y')\in ]0,W[\times]W,+\infty[\ \ \ , &(2.2{\rm b})
\cr\cr
\beta \phiphi&=-\ln\left|{z-\bar{z}'-iW\over z-\bar{z}'}\right|
 \cr
 &\hbox{if}\quad (y,y')\in ]W,+\infty[^2 \ \ \ .&(2.2{\rm c})
\cr
}
$$
\noindent In the region $y \leq 0$, $\Phi ({\bf r}) = 0$ without fluctuations.
\par

 From (2.2), one can compute the correlations of the electric field ${\bf
E}({\bf r})$,
and from the discontinuities of $E_y({\bf r})$ on the lines $y = 0$ and $y = W$
one
obtains the correlations for the charge densities $\sigma({\bf r})$ (charge per
unit
length) on these lines~:
$$
\eqalignno{
\beta \sigmasigma&=-{1\over 2\pi^2}\left[
\left(\piw\right)^2\left({1\over\sinh\piw (x-x')}\right)^2 -
\left({1\over x-x'}\right)^2\right]
\cr
&\hbox{if {\bf r} and {\bf r}' are on the ideal conductor,}&(2.3{\rm a})
\cr
}$$
$$\eqalignno{
\beta &\sigmasigma =-{1\over 2}
\left({1\over 2W}\right)^2\left({1\over\cosh\piw (x-x')}\right)^2
\cr
&\hbox{if {\bf r} is on the ideal conductor and {\bf r}' on the
Coulomb system,}&(2.3{\rm b})
\cr
}$$
\noindent and
$$\eqalignno{
\beta \sigmasigma&=-{1\over 2\pi^2}\left[
\left(\piw\right)^2\left({1\over\sinh\piw (x-x')}\right)^2 +
\left({1\over x-x'}\right)^2\right]
\cr
&\hbox{if {\bf r} and {\bf r}' are on the Coulomb system.}&(2.3{\rm c})
\cr
}
$$
\noindent [(2.3b) and (2.3c) disregard the microscopic detail~; (2.3c) must be
regularized at $x - x' =0$]. As expected, if $W \to \infty$, the fluctuations
(2.3a) on
the ideal conductor $C$ now alone in space disappear, as well as the
correlations (2.3b)
between the ideal conductor $C$ and the Coulomb system $S$, while the
fluctuations (2.3c)
on the Coulomb system $S$ become those of one conducting line [6]. \par

 From the averages $<E_{\mu}E_{\nu}>$ associated to (2.2), one can compute the
Maxwell
stress tensor. Its only non-zero component is $T_{yy}$. At any point ${\bf r}$
between
the lines $y = 0$ and $y = W$, assuming that there is no potential difference
between
the lines, one found [7]
$$
\beta T_{yy}={\pi\over 24W^2}\ \ \ . \eqno(2.4)
$$
\noindent $- T_{yy}$ is the force per unit length exerted on the Coulomb system
$S$. \par \medskip

\vskip 5mm
\noindent {\it 2.3 The high-density limit approach}\vskip 5mm

We now assume that both lines $y=0$ and $y=W$ are Coulomb systems, and
the high-density limit is taken on the line $y=0$. In that limit, the
macroscopic approach [6] for computing the potential correlations is
valid under the same conditions as in section~2.2~: {\bf r} and {\bf r}'
should be at macroscopic distances from the line $y=W$, but there is no
restriction about their distances to the line $y=0$. One finds
$$
\eqalignno{
\beta \phiphi&=-\ln\left| z-\bar{z}'\right|
 \cr
&\hbox{if} \quad(y,y')\in ]-\infty,0[^2 \ \ \ , &(2.5{\rm a})
\cr
}$$
$$\eqalignno{
\beta \phiphi&=-\ln\left| (z-z'){{\sinh\piw (z-\bar{z}')}
\over {\sinh\piw (z-z')}}\right| \cr
& \hbox{if}\quad (y,y')\in ]0,W[^2 \ \ \ ,&(2.5{\rm b})
\cr
}$$
$$\eqalignno{
\beta \phiphi&=-\ln\left| z-z'\right|
 \cr
& \hbox{if} \quad(y,y')\in ]0,W[\times]W,+\infty[ \ \ \ , &(2.5{\rm c})
\cr
}$$
$$\eqalignno{
\beta \phiphi&=-\ln\left| z-\bar{z}'-iW \right|
 \cr
 & \hbox{if}\quad (y,y')\in ]W,+\infty[^2 \ \ \ , &(2.5{\rm d})
\cr
}$$
$$\eqalignno{
\beta \phiphi&=-\ln\left| z-z'\right|
 \cr
 &\hbox{if} \quad(y,y')\in ]-\infty,0[\times]W,+\infty[ \ \ \ . &(2.5{\rm e})
\cr
}
$$
The difference between the electric potential correlations in this case
and the previous case [equations~(2.2)] is the electric potential
correlation of a system where there is only one
Coulomb system on the line $y=0$ [6]. This difference raises from
neglecting the fluctuations of the conductor at $y=0$ in the ideal
conductor case.

 From (2.5) we obtain the charge correlations
$$
\eqalignno{
\beta \sigmasigma&=-{1\over 2\pi^2}\left[
\left(\piw\right)^2\left({1\over\sinh\piw (x-x')}\right)^2 +
\left({1\over x-x'}\right)^2\right]\cr
& \hbox{if}
\quad y=y'=0\ \ \ , &(2.6{\rm a})
\cr\cr
\beta \sigmasigma&=-{1\over 2}
\left({1\over 2W}\right)^2\left({1\over\cosh\piw (x-x')}\right)^2
\cr
 & \hbox{if}\quad y=0\quad\hbox{and}\quad y'=W \ \ \ , &(2.6{\rm b})
\cr\cr
\beta \sigmasigma&=-{1\over 2\pi^2}\left[
\left(\piw\right)^2\left({1\over\sinh\piw (x-x')}\right)^2 +
\left({1\over x-x'}\right)^2\right] \cr
&\hbox{if}\quad y=y'=W  &(2.6{\rm c})
\cr\cr
}
$$
\noindent [eqs. (2.6) disregard the microscopic detail~; (2.6a) and (2.6b) must
be
regularized at $x - x' = 0$]. \par

The charge correlations in the Coulomb system at $y=W$
are the same in both approaches [equations~(2.3c) and (2.6c)] as it was noticed
before in
other models~[4, 2]. Also the correlation between a point on the conductor
($y=0$) and a point on the Coulomb system ($y=W$) is the same in both
approaches [equations~(2.3b) and (2.6b)]. But the
correlation between two points on the conductor at $y=0$ differ when the
conductor is ideal [equation~(2.3a)] and when it is the high-density limit of a
Coulomb
system [equation~(2.6a)].

The stress tensor is the same as in equation~(2.4).
This shows that we have the same results in both approaches for the
force exerted on the Coulomb system S. What is special to the present model and
what is
general will be discussed in section~3. \vskip 5mm

\noindent {\it 2.4. A solvable model} \vskip 5mm
The charge correlations (2.3) and (2.6) can be checked on a solvable
microscopic
model. In this section we consider that the Coulomb system $S$ is a
one-component
plasma~: the system is composed of particles of charge $q$ moving in a rigid
charged
background. The two-dimensional one-component plasma is a solvable
model in several geometries [8-11] when $\beta q^2=2$. \par

a) The system such that $C$ is an ideal conductor has been solved in [12, 13,
7], in the
grand canonical ensemble. Let $-q \eta$ be the background charge density of $S$
and
$\zeta$ the fugacity. The number density $n$ and charge correlation in the
Coulomb system
$S$ are given in terms of
$$
g(x)=\int_0^\infty {dk\over 2\pi}{e^{ikx}\over 1+(2\pi\zeta)^{-1}
e^{2W(k-2\pi\eta)}}
\eqno(2.7)
$$
as
$$
n=g(0)
\eqno(2.8)
$$
and
$$
\sigmasigma=-q^2|g(x-x')|^2+q^2n\delta(x) \ \ \ .
\eqno(2.9)
$$

It has been shown in [7] that, in the macroscopic limit ($\eta W>>1$), these
results agree with those from the macroscopic approach of section 2.2~; (2.9)
becomes
(2.3a). \par

b) The system such that $C$ is a conducting Coulomb system can also be solved
exactly. Now each line is a one-component plasma. The background charge
densities are
$-q\eta_0$ for the line $y=0$ and $-q \eta$ for the line $y=W$. Here we work in
the
canonical ensemble. Let N be the total number of particles. We consider first
that we have
two concentric circles, with radii $R$ and $R + W$, on which each plasma lies,
and then
take the limit $R \to \infty$, with $N = 2 \pi \eta_0 R + 2 \pi \eta (R + W)$,
which ensures the overall neutrality. Adapting [11] by treating the radial
coordinate $r$
as a discrete variable which can have the values $R$ and $R + W$, we introduce
the $N$
orthogonal functions
$$\psi_{\ell}({\bf r}) = \left ( \alpha \ \delta_{r,R} + \delta_{r,
r+W} \right ) z^{\ell} \quad , \qquad 0 \leq \ell \leq N-1 \ \ \ ,
\eqno(2.10)$$
\noindent where $z = r\ e^{i\theta}$, $\delta$ is the Kronecker symbol, and
$\alpha$ is a
(positive) parameter which controls [14] how the $N$ particles are distributed
between
the two lines. The density $n$ and charge correlations are given in terms of
the
projector
$$P(\rrprime)=\sum_{\ell=0}^{N-1} {\Psi_\ell({\bf r})\overline{\Psi_\ell({\bf
r}')}
\over
\sum_{r_0\in\{R,R+W\}}\int_0^{2\pi}
 |\Psi_\ell(r_0,\theta_0)|^2 r_0 d{\theta_0}}
\eqno(2.11)
$$
as
$$
n({\bf r})=P({\bf r},{\bf r})\eqno(2.12)
$$
and
$$
\sigmasigma = -q^2 |P(\rrprime)|^2 + q^2 n({\bf r}) \delta_{r,r'} \delta
(r(\theta -
\theta ')) \ \ \ .\eqno(2.13) $$

\noindent In the limit $R\to\infty$ the two circles become two parallel lines.
In
this limit it is useful to define $k=\ell / R$. A summation over $\ell$
becomes an integral over $k$ times $R$. We change our system of
coordinates: let $x=R\theta$ and $y=r-R$. The projector becomes
$$
\eqalignno{
P(\rrprime)=&
\delta_{y,y'}(\delta_{y,0}P_1(x-x')+\delta_{y,W}P_2(x-x'))
\cr
&+(\delta_{y,0}\delta_{y',W}+\delta_{y,W}\delta_{y',0})
P_3(x-x')
&(2.14)
}
$$
with
$$P_1(x) = {1 \over 2 \pi} \int_0^{2 \pi (\eta_0 + \eta )} {e^{ikx} dk \over 1
+
\alpha^{-2} \ e^{2Wk}} \ \ \ , \eqno(2.15{\rm a})$$

$$P_2(x) = {1 \over 2 \pi} \int_0^{2 \pi (\eta_0 + \eta )} {e^{ikx} dk \over 1
+ \alpha^2
e^{-2Wk}} \ \ \ , \eqno(2.15{\rm b})$$

$$P_3(x) = {1 \over 2 \pi} \int_0^{2 \pi (\eta_0 + \eta )} {e^{ikx} dk \over
\alpha \
e^{-Wk} + \alpha^{-1} e^{Wk}} \ \ \ . \eqno(2.15{\rm c})$$
\noindent For a comparison with previous results, we define an alternative
control
parameter $\zeta$ by
$$2 \pi \zeta = \alpha^{-2} e^{4 \pi \eta_0W} \eqno(2.16)$$
\noindent and keep $\zeta$ fixed as we vary the other parameters $\eta_0$,
$\eta$, $W$.
\par

When the density $\eta_0$ of the conductor $C$ becomes infinite, from (2.15b)
where we
make the change of variable $k \to 2 \pi (\eta_0 + \eta ) - k$, we obtain
$$\left | P_2(x) \right | = \left | {1 \over 2 \pi} \int_0^{\infty} {e^{-ikx}
dk \over 1
+ (2 \pi \zeta )^{-1} \ e^{2W(k-2 \pi \eta)}} \right | \ \ \ . \eqno(2.17)$$

\noindent Since $|P_2(x)|$ as given by (2.17) is identical to $|g(x)|$ as given
by
(2.7), the density and the charge correlation function on the Coulomb system
$S$ are
indeed identical whenever $C$ is an ideal conductor or the high-density limit
of a
Coulomb system. \par

A more detailed comparison can be made when $W$ is macroscopic. Then,
neglecting in
(2.15) terms $\exp (- \eta_0 W)$ and $\exp (- \eta W)$, after an averaging over
oscillations of microscopic wavelength we obtain
$$
|P_1(x)|^2 \sim |P_2(x)|^2
\sim {1 \over 4 \pi^2} \left [ \left({1\over x}\right)^2+\left({\pi\over
2W\sinh{\pi
x\over 2W}}\right)^2 \right ] \ \ \ , \eqno(2.18{\rm a})
$$
$$
|P_3(x)|^2\sim {1 \over 4 \pi^2} \left({\pi\over 2W\cosh{\pi x\over
2W}}\right)^2 \ \ \ .
\eqno(2.18{\rm b})
$$
The charge correlations obtained from (2.18) agree with the macroscopic ones
obtained in
(2.6).

\vfill \supereject
\noindent {\bf 3. General case}\vskip 5mm

In this section we consider the general case in $d\geq 2$ dimensions. The
conductor
$C$ has any shape and the Coulomb system $S$ occupies some region of space
outside $C$.
The Coulomb potential is
$$
G_0({\bf r})=\cases{
-\ln r, & if $d=2$\cr
r^{2-d}, & if $d>2$ \ \ \ .\cr
}\eqno(3.1)
$$
\noindent To start with, $C$ itself is considered as a Coulomb system with
internal
degrees of freedom. \par

We shall use several quantities related to the electric potential
correlations for the conductor $C$ alone in space. Let $\Phi_{\rm c}({\bf r})$
be the
electric potential at ${\bf r}$ created by the conductor $C$ alone and let
$<...>_0^{\rm T}$ be a truncated statistical average computed with the
Boltzmann weight of
the conductor $C$ alone. The correlation $\correl{\Phi_{\rm c}}{\Phi_{\rm
c}}_0$ can be
computed by linear response [6]~; it is related to the average electric
potential change
at ${\bf r}$ when a unit charge is put at ${\bf r}'$. This potential change can
be
computed by macroscopic electrostatics. If the conductor is grounded, there are
two
cases~: \par
a) If {\bf r} (or {\bf r}', or both) is (are) inside the conductor,
$$\beta <\Phi_{\rm c}({\bf r}) \Phi_{\rm c}({\bf r}')>_0^{\rm T}=G_0({\bf r} -
{\bf r}') \
\ \ .  \eqno(3.2{\rm a})
$$ \par

b) If {\bf r} and {\bf r}' are outside the conductor,
$$ \beta <\Phi_{\rm c}({\bf r}) \Phi_{\rm c}({\bf r}')>_0^{\rm T} =G^*({\bf r},
{\bf r}')
\ \ \ ,  \eqno(3.2{\rm b})
$$
\noindent where $G^*$ is defined by
$$
\Delta_{\bf r}
\left[G_0({\bf r}-{\bf r}')-G^*({\bf r},{\bf r}')\right]
=
-\mu_d\delta({\bf r}-{\bf r}')\eqno(3.3)
$$
for {\bf r} and {\bf r}'
outside the conductor,
with $\mu_2=2\pi$, $\mu_3 = 4 \pi$, ...  $\mu_d=(d-2)2\pi^{d/2}/\Gamma(d/2)$ if
$d>2$,
and the condition $G_0({\bf r}-{\bf r}')-G^*({\bf r},{\bf r}')=0$
if {\bf r} (or {\bf r}') is on the surface of the conductor~; $G_0({\bf r}-{\bf
r}')-G^*({\bf r},{\bf r}')$ is the electric potential at {\bf r} created by a
unit charge at {\bf r}' in presence of a grounded ideal conductor. The
expressions (3.2)
which disregard the microscopic detail become exact in the limit when the
correlation
length goes to zero. \par

Another remark useful for the following sections
is that the fluctuations of the electric potential of a conductor $C$ are
Gaussian [6,
15, 16].

Let ${\bf R}$ be the set of particle coordinates of the conductor $C$,
$\{{\bf r}_i\}$ the set of the $N$ particle coordinates of the Coulomb system
$S$,
$d\Gamma$ the element of phase space of the Coulomb system
and $H_0({\bf R})$ the Hamiltonian of the conductor. The total energy of the
system (S plus C) is
$$
H(\{{\bf r}_i\},{\bf R})=\sum_{i=0}^N q_i \Phi_{\rm c}({\bf r}_i, {\bf R}) +
\sum_{1\leq i<j\leq N}  q_i q_j G_0({\bf r}_i -{\bf r}_j)+
H_0(\bf R)\ \ \ , \eqno(3.4)
$$
where ${\bf r}_i$ and $q_i$ are the position and charge of the $i^{\rm
th}$ particle of $S$. There might also be some short-range interaction between
the
particles of $S$, but we do not explicitly write it in (3.4) just for having a
simpler
notation.

In the following sections we shall compute the partition function, the
correlations, the
stress tensor and the force exerted on the Coulomb system $S$, in the limit
when the
charge correlation length of the conductor $C$ goes to zero (we shall call this
limit
the good conductor limit)  and compare the results to the ones
when the conductor is ideal (i.e. the potential on it is fixed, say to zero,
without
fluctuations). For the sake of simplicity we shall only treat in detail this
case of a
grounded conductor, but similar results hold for an insulated conductor (see
section
3.5).

\vskip 5mm
\vbox{
\noindent {\it 3.1 Partition function and statistical averages}\vskip 5mm

The partition function of the total system can be written as
}
$$
Z=\int {d\Gamma}
<e^{-\beta\sum_i q_i \Phi_{\rm c}({\bf r}_i)}>_0
e^{-\beta\sum_{i<j} q_i q_j G_0({\bf r}_i-{\bf r}_j)
}
Z_0\eqno(3.5)
$$
where $<...>_0$ means the average over ${\bf R}$ with the Boltzmann
weight $\exp(-\beta H_0({\bf R}))$ and
$Z_0=\int\exp(-\beta H_0({\bf R}))d{\bf R}$ is
the partition
function of the conductor $C$ alone.

Now, since the fluctuations of $\Phi_{\rm c}$ are Gaussian,
$$
\eqalignno{
<\exp\left[-\beta\sum_i q_i \Phi_{\rm c}({\bf r}_i)\right]>_0
&=
\exp\left[ {1\over 2}\beta^2 \sum_{i,j} q_i q_j
<\Phi_{\rm c}({\bf r}_i) \Phi_{\rm c}({\bf r}_j)>^{\rm T}_0\right]
\cr
&=
\exp\left[{1\over 2}\beta\sum_{i,j} q_i q_j
G^*({\bf r}_i,{\bf r}_j)
\right] \ \ \ , &(3.6)
\cr
}
$$
where we have used (3.2b). Thus, the partition function becomes
$$
Z=Z^*Z_0\eqno(3.7)
$$
where $Z^*$ is the partition function of the Coulomb system $S$ in presence
of an ideal conductor,
$$Z^* = \int d\Gamma \ e^{-\beta H_{\rm eff}} \eqno(3.8)$$
\noindent where
$$H_{\rm eff}({\bf r}_1,..,{\bf r}_N)=
 -{1\over 2} \sum_{i=1}^N
 q_i^2 G^*({\bf r}_i,{\bf r}_i)
+ \sum_{1\leq i<j\leq N}  q_i q_j
\left[ G_0({\bf r}_i -{\bf r}_j)-
G^*({\bf r}_i,{\bf r}_j)\right] \ \ \ . \eqno(3.9)
$$
\noindent $H_{\rm eff}$ is indeed the standard Hamiltonian used in the ideal
conductor
approach. For instance, in the case of a plane ideal conductor, $G^*$ is the
particle-image interaction~; it should be noted that the interaction $-q_i^2
G^*({\bf
r}_i, {\bf r}_i)$ of a particle with its own image carries a factor 1/2 in
(3.9). \par

The total free energy is $F=F^*+F_0$ where $F^*$ is the free energy of $S$ in
the
presence of an ideal conductor and $F_0$ the free energy of the conductor $C$
alone. This
was noticed previously in [4] for the model of a two-dimensional plasma near a
metallic wall.

Let $A(\{{\bf r}_i\})$ be a microscopic quantity that does not depend on {\bf
R}. Its
thermodynamic average is
$$
\eqalignno{
<A>&={1\over Z^*}\int
{d\Gamma}
<e^{-\beta\sum_i q_i \Phi_{\rm c}({\bf r}_i)}>_0
e^{-\beta\sum_{i<j} q_i q_j G_0({\bf r}_i-{\bf r}_j)
}A\cr
&={1\over Z^*}\int
{d\Gamma}
e^{-\beta H_{\rm eff}({\bf r}_1,..,{\bf r}_N)}A
\cr
&=<A>_{\rm eff} &(3.10)\cr
}
$$
\noindent where we have used (3.6). Thus the average of A can be
computed by assuming from the beginning that the conductor $C$ is ideal.

\vskip 5 truemm
\noindent {\it 3.2 Electric potential correlations}
\vskip 5mm

Equation (3.10) does not apply to the electric potential
correlations because the mi\-cros\-copic electric potential is
different in the cases of a good conductor or an ideal conductor. For the
good conductor case, the microscopic electric potential is

$$
\Phi({\bf r})=
\sum_iq_iG_0({\bf r}-{\bf r}_i) + \Phi_{\rm c}({\bf r}) \ \ \ ,
\eqno(3.11)
$$
while for the ideal conductor case it is
$$\Phi_{\rm id}({\bf r})= \left \{ \matrix{
0, \quad \hbox{if} \ {\bf r} \in C \hfill \cr
\cr
\sum_iq_i\left[G_0({\bf r}-{\bf r}_i)-G^*({\bf r},{\bf r}_i)\right], \quad
\hbox{if}
\ {\bf r} \notin C\ \ \ . &\hskip 4.6 truecm (3.12) \cr
} \right .
$$

The average electric potential in the good conductor case is
$$
\eqalignno{
<\Phi({\bf r})>=
{1\over Z^*} \int {d\Gamma}& \Big \{
\sum_iq_iG_0({\bf r}-{\bf r}_i)e^{-\beta H_{\rm eff}}
\cr
&
+<\Phi_{\rm c}({\bf r})e^{-\beta\sum_i q_i\Phi_{\rm c}({\bf r}_i)}>_0
e^{-\beta\sum_{i<j}q_iq_jG_0({\bf r}_i-{\bf r}_j)} \Big \} \ \ \ . &(3.13) \cr
}
$$
\noindent Since the fluctuations of $\Phi_{\rm c}$ are Gaussian,
$$
<\Phi_{\rm c}({\bf r})
e^{-\beta\sum_iq_i\Phi_{\rm c}({\bf r}_i)}>_0
=
- \beta \sum_iq_i<\Phi_{\rm c}({\bf r}) \Phi_{\rm c}({\bf r}_i)>_0
e^{{1\over 2}\beta^2 \sum_{i,j}q_iq_j<\Phi_{\rm c}({\bf r}_i)\Phi_{\rm c}({\bf
r}_j)>_0} \
\ \ . \eqno(3.14)
$$
\noindent Using the covariance (3.2), one finds, for all ${\bf r}$,
$$
<\Phi({\bf r})>=<\Phi_{\rm id}({\bf r})>_{\rm eff}
\eqno(3.15)
$$
where $<\Phi_{\rm id}({\bf r})>_{\rm eff}$ is the average electric
potential in the ideal conductor case.

We can compute the electric potential correlations in the same way~:
the correlation function in the good conductor case can be written as
$$
\eqalignno{
\beta <\Phi({\bf r})\Phi({\bf r}')>
 = {1\over Z^*} &\int {d\Gamma} \Big \{
\sum_{i,j} q_iq_jG_0({\bf r}-{\bf r}_i)G_0({\bf r}'-{\bf r}_j)
e^{-\beta H_{\rm eff}}
\cr
&+e^{-\beta\sum_{i<j}q_iq_jG_0({\bf r}_i-{\bf r}_j)}
\cr
&\times\Big[
\sum_iq_iG_0({\bf r}-{\bf r}_i)<\Phi_{\rm c}({\bf r}')
e^{-\beta\sum_iq_i\Phi_{\rm c}({\bf r}_i)}>_0
\cr
&+\sum_iq_iG_0({\bf r}'-{\bf r}_i)<\Phi_{\rm c}({\bf r})
e^{-\beta\sum_iq_i\Phi_{\rm c}({\bf r}_i)}>_0
\cr
&+<\Phi_{\rm c}({\bf r})\Phi_{\rm c}({\bf r}')
e^{-\beta\sum_iq_i\Phi_{\rm c}({\bf r}_i)}>_0
\Big] \Big \} \ \ \ . &(3.16)\cr
}
$$
Using (3.15),
$$
\eqalignno{
<\Phi_{\rm c}({\bf r})\Phi_{\rm c}({\bf r}')
e^{-\beta\sum_iq_i\Phi_{\rm c}({\bf r}_i)}>_0=&
\Big[ \beta^2
\sum_{i,j}q_iq_j<\Phi_{\rm c}({\bf r}) \Phi_{\rm c}({\bf r}_i)>_0 <\Phi_{\rm
c}({\bf r}')\Phi_{\rm c}({\bf
r}_j)>_0  \cr
&+<\Phi_{\rm c}({\bf r}) \Phi_{\rm c}({\bf r}')>_0
\Big ]\cr
&\times e^{{1\over 2}\beta^2 \sum_{i,j}q_iq_j<\Phi_c({\bf r}_i)\Phi_c({\bf
r}_j)>_0}
&(3.17) \cr
}
$$
\noindent [also a consequence of $\Phi_{\rm c}$ being Gaussian], and the
covariance
(3.2), we find $$
\phiphi={\correl{\Phi_{\rm id}}{\Phi_{\rm id}}}_{\rm eff}
+<\Phi_{\rm c}({\bf r}) \Phi_{\rm c}({\bf r}')>_0^{\rm T} \ \ \ .
\eqno(3.18)
$$

Thus, the correlation function in presence of a good conductor is the
correlation
function in presence of an ideal conductor plus the correlation function for
the good
conductor alone in space.

This is what was noticed in the example of section 2.

\vskip 5mm
\noindent {\it 3.2 Charge correlations}
\vskip 5mm
If we are interested in charge correlations in the Coulomb system $S$,
equation (3.10) applies because the microscopic charge density outside
the conductor
$$
\rho({\bf r})=\sum_{i=1}^N q_i\delta({\bf r}-{\bf r}_i)
\eqno(3.19)
$$
does not depend on the coordinates {\bf R}~; thus the charge correlations
inside $S$ are
the same in both approaches. The surface charge density on $C$ is given by the
discontinuity of the normal electric field, thus using (3.15) we find that the
average
charge density is the same in both approaches. The same holds for the
correlation
between the density in $S$ and the surface charge density on $C$. But, if we
are
interested in charge correlations on the conductor $C$, the correlations are
different
in the two approaches. Using equation (3.18) we can compute the difference in
the
electric field correlations and from it the difference in the charge
correlations on the
surface of $C$~; this difference is the surface charge correlation on $C$ when
it is
alone in space. \par

The example of section 2 illustrates these general results.

\vskip 5mm
\noindent {\it 3.4 The stress tensor and the forces exerted on the
Coulomb system}
\vskip 5mm

The Maxwell stress tensor is
$$
T_{\mu\nu}=\mu_d^{-1}
<E_\mu E_\nu - {\delta_{\mu\nu}\over 2}{\bf E}^2
>
\eqno(3.20)
$$
where ${\bf E}=-\nabla\Phi$ is the electric field. Let ${\cal V}$ be some
volume
outside the conductor $C$. The total average electric force on ${\cal V}$ is
$${\bf F}=\int_{\partial{\cal V}} {\bf T}.d^{d-1}{\bf S}=
\int_{\cal V} \nabla . {\bf T}({\bf r}) d^d{\bf r} \ \ \ .
\eqno(3.21)
$$

It can be shown that this force is the same in both models although the
electric potential correlations are different and consequently the
stress tensor might be different. Indeed, from equation~(3.18), the difference
between the
stress tensor in the good conductor case and the ideal conductor case is
$$
T_{\mu\nu}({\bf r})-T_{\mu\nu}^{\rm id}({\bf r})
=
\mu_d^{-1} \left(\partial_\mu \partial_\nu' - {\delta_{\mu\nu}\over 2}
\partial_\sigma\partial_\sigma'
\right)G^*({\bf r},{\bf r})
\eqno(3.22)
$$
where $\partial$ (resp. $\partial'$) means partial differentiation
with respect to the first (resp. the second) argument of $G^*$.
The difference of the divergences is
$$
\partial_\mu\left[
T_{\mu\nu}({\bf r})-T_{\mu\nu}^{\rm id}({\bf r})
\right]
=\mu_d^{-1}
\left(
\partial_\nu'\partial_\mu\partial_\mu G^*({\bf r},{\bf r})
+{1\over 2}
\partial_\mu\partial_\mu'\left[\partial_\nu'G^*({\bf r},{\bf r})
-\partial_\nu G^*({\bf r},{\bf r})\right]
\right) \ \ \ .
\eqno(3.23)
$$
Now, from~(3.3) $\partial_\mu\partial_\mu G^*({\bf
r},{\bf r})=0$, and since $G^*$ is
symetrical $\partial_\nu'G^*({\bf r},{\bf r})
=\partial_\nu G^*({\bf r},{\bf r})
$. Thus, (3.23) is zero and the force is the same in both models. \par

In the example of section~2, the stress tensor itself was the same in presence
of either
a good conductor or an ideal conductor. But this was an effect of the very
peculiar
symmetry of the model (invariance by translations along the $x$ axis). \par

A direct evaluation of the average force exerted on one particle of the system
confirms
that the two approaches give the same result. Let $d\Gamma_{N-1}$ be the phase
space
element of the system $S$ when the $i^{\rm th}$ particle is fixed and
$Z^*_{N-1}$ the
partition function of the system $S$ in presence of an ideal conductor when the
$i^{\rm
th}$ particle is fixed. The average force exerted on the $i^{\rm th}$ particle
in
presence of a good conductor is
$$
\eqalignno{
<{\bf F}_i> =-&{1\over Z^*} \int d\Gamma_{N-1}\Bigg\{
\nabla_i\left[\sum_{l<k}q_lq_kG_0({\bf r}_l-{\bf r}_k)\right]
e^{-\beta H_{\rm eff}}
\cr
&+<\nabla_i\left[\sum_lq_l\Phi_{\rm c}({\rm r}_l)\right]
e^{-\beta\sum_kq_k\Phi_{\rm c}({\rm r}_k)}>_0
e^{-\beta\sum_{l<k}q_lq_kG_0({\bf r}_l-{\bf r}_k)}
\Bigg\}. &(3.24)
\cr
}
$$
In the last term
$$
\eqalign{
<\nabla_i\left[\sum_lq_l\Phi_{\rm c}({\rm r}_l)\right]
e^{-\beta\sum_kq_k\Phi_{\rm c}({\rm r}_k)}>_0
&=
-\beta^{-1}\nabla_i\left[
<e^{-\beta\sum_kq_k\Phi_{\rm c}({\rm r}_k)}>_0
\right]
\cr
&=
-\beta^{-1}\nabla_i\left[
e^{{\beta\over 2}\sum_{kl}q_kq_l G^*({\bf r}_k,{\bf r}_l)}
\right]
\cr
&=
-\nabla_i\left[\sum_{kl}{q_kq_l\over 2} G^*({\bf r}_k,{\bf r}_l)
\right]
e^{{\beta\over 2}\sum_{kl}q_kq_l G^*({\bf r}_k,{\bf r}_l)} \ \ .
\cr
}
\eqno(3.25)
$$
Using (3.25) in~(3.24) gives
$$
<{\bf F}_i>=<{\bf F}^{\rm id}_i>_{\rm eff} \ \ \ .
\eqno(3.26)
$$
We obtain the same average force by both approaches.

\vskip 5mm
\noindent {\it 3.5 The insulated conductor case}
\vskip 5mm

In the former calculations we assumed that the conductor $C$ was grounded.
The same calculations can be carried out if the conductor $C$ is insulated.
Equations
(3.7), (3.10), (3.15), (3.26) still hold
for the insulated conductor case. The expression of $H_{\rm eff}$ now is a
different
one but it is still the Hamiltonian of the system $S$ in presence of an ideal
conductor~: everywhere $G^*({\bf r}, {\bf r}')$ must be replaced by $G^*({\bf
r}, {\bf
r}') + Q({\bf r}) \ V({\bf r}')$ where $Q({\bf r})$ is the charge created by
influence
of a unit charge at ${\bf r}$ on the grounded conductor and $V({\bf r}')$ is
the
electric potential at ${\bf r}'$ created by the conductor carrying a
unit charge. The important fact to notice is that the relation (3.18) between
the
different potential correlation functions is still valid.

\vskip 5mm
\noindent {\bf 4. Conclusion}
\vskip 5mm

Two different methods for treating the problem of a Coulomb system near a
conductor have been compared. \par

The first method, where the conductor is considered
from the beginning as ideal, neglects all fluctuations in the conductor.
The second method treats the conductor as a conducting Coulomb system the
charge
correlation length of which goes to zero. Even in that limit, the fluctuations
in the
conductor do not vanish. This modifies the electric potential correlations by a
term
given by the electric potential correlation when the conductor is alone, but
the
average electric potential is not modified. Because of this, quantities such as
the
electric field and charge density will have the same properties~: their average
is
the same in both methods but their correlations differ by the correlation for
the
conductor alone. However, in the case of the charge correlations, this
difference
vanishes outside the conductor. \par

The free energy in the good conductor case is just
the sum of the free energy in the ideal conductor case plus the free energy
when the
conductor is alone in space. \par

The average stress tensor is modified only by a term the divergence of which is
zero,
so the average force exerted on any part of the Coulomb system is not modified.
However,
the fluctuations of these forces will in general be different in the two
methods.

\vfill \supereject
\noindent {\bf References} \par
  \vskip 5mm

\item{[1]} Forrester P~J 1985 {\it J.~Phys.~A: Math.~Gen.} {\bf 18} 1419

\item{[2]} Forrester P~J 1991 {\it J.~Chem.~Phys.} {\bf 95} 4545

\item{[3]} Forrester P~J 1992 {\it J.~Stat.~Phys.} {\bf 67} 433

\item{[4]} Alastuey A, Jancovici B, Blum L, Forrester P~J and Rosinberg
M~L 1985 {\it J.~Chem.~Phys.} {\bf 83} 2366

\item{[5]} Cornu F and Jancovici B 1989 {\it J.~Chem.~Phys.} {\bf 90}
2444

\item{[6]} Jancovici B 1995 {\it J. Stat. Phys.} {\bf 80} 455

\item{[7]} Forrester P J, Jancovici B, and T\'ellez G 1995 {\it Universality in
Some
Classical Coulomb Systems of Restricted Dimension} LPTHE Orsay Preprint 95-55

\item{[8]} Jancovici B 1981 {\it Phys. Rev. Lett.} {\bf 46} 386

\item{[9]} Caillol JM 1981 {\it J.~Phys. - LETTRES (Paris)} {\bf 42} L-245

\item{[10]} Alastuey A and Lebowitz J~L 1984 {\it J.~Phys.~(Paris)} {\bf
45} 1859

\item{[11]} Cornu F, Jancovici B and Blum L 1988 {\it J.~Stat.~Phys} {\bf
50} 1221

\item{[12]} Gaudin M 1966 {\it
Nucl. Phys.} {\bf 85} 545

\item{[13]} Forrester PJ 1993 {\it
Phys. Lett. A} {\bf 173} 355

\item{[14]} Rosinberg ML and Blum L 1984 {\it
J. Chem. Phys.} {\bf 81} 3700

\item{[15]} Jancovici B, Manificat G and Pisani C 1994 {\it
J.~Stat.~Phys.} {\bf 76} 307

\item{[16]} Jancovici B and T\'ellez G 1996 {\it J.~Stat.~Phys.} {\bf
82} 609.

\bye